\documentclass{aastex63}

\usepackage{graphicx,,epsfig,fancyhdr,rotating,amsmath,epsf, txfonts,natbib,epstopdf,multirow,booktabs,longtable,color,hyperref}

\received{***}
\revised{***}
\accepted{***}
\submitjournal{ApJL}

\shorttitle{Quasi-periodic EUV Wave Train}
\shortauthors{Sun et al.}
\graphicspath{{./}{figures/}}

\begin{document}

\title{Cross-loop propagation of a quasi-periodic extreme-ultraviolet
wave train triggered by successive stretching of magnetic field structures during a solar eruption}

\author[0000-0001-5657-7587]{Zheng Sun}
\affiliation{School of Geophysics and Information Technology, China University of Geosciences (Beijing) Beijing, 100083, People's Republic of China; \href{yaoshuo@cugb.edu.cn}{yaoshuo@cugb.edu.cn}}

\author[0000-0002-1369-1758]{Hui Tian}
\affiliation{School of Earth and Space Sciences, Peking University, Beijing, 100871, People's Republic of China; \href{huitian@pku.edu.cn}{huitian@pku.edu.cn}}
\affiliation{Key Laboratory of Solar Activity and Space Weather, National Space Science Center, Chinese Academy of Sciences, Beijing 100190, People's Republic of China}

\author[0000-0002-7289-642X]{P. F. Chen}
\affiliation{School of Astronomy and Space Science, Nanjing University, Nanjing 210023, People's Republic of China}
\affiliation{Key Laboratory of Modern Astronomy and Astrophysics (Nanjing University), Ministry of Education, Nanjing 210023, People's Republic of China}

\author[0000-0003-4267-0486]{Shuo Yao}
\affiliation{School of Geophysics and Information Technology, China University of Geosciences (Beijing) Beijing, 100083, People's Republic of China; \href{yaoshuo@cugb.edu.cn}{yaoshuo@cugb.edu.cn}}

\author[0000-0003-4804-5673]{Zhenyong Hou}
\affiliation{School of Earth and Space Sciences, Peking University, Beijing, 100871, People's Republic of China; \href{huitian@pku.edu.cn}{huitian@pku.edu.cn}}
\affiliation{Key Laboratory of Modern Astronomy and Astrophysics (Nanjing University), Ministry of Education, Nanjing 210093, People's Republic of China}

\author[0000-0001-7866-4358]{Hechao Chen}
\affiliation{School of Earth and Space Sciences, Peking University, Beijing, 100871, People's Republic of China; \href{huitian@pku.edu.cn}{huitian@pku.edu.cn}}

\author[0000-0002-3618-3430]{Linjie Chen}
\affiliation{Key Laboratory of Solar Activity and Space Weather, National Space Science Center, Chinese Academy of Sciences, Beijing 100190, People's Republic of China}

\begin{abstract}
Solar extreme-ultraviolet (EUV) waves generally refer to large-scale disturbances propagating outward from sites of solar eruptions in EUV imaging observations. Using the recent observations from the Atmospheric Imaging Assembly (AIA) on board the Solar Dynamics Observatory (SDO), we report a quasi-periodic wave train propagating outward at an average speed of $\sim$308 km s$^{-1}$. At least five wavefronts can be clearly identified with the period being $\sim$120 s. These wavefronts originate from the coronal loop expansion, which propagates with an apparent speed of $\sim$95 km s$^{-1}$, about 3 times slower than the wave train. In the absence of a strong lateral expansion, these observational results might be explained by the theoretical model of \citet{chen2002evidence}, which predicted that EUV waves may have two components: a faster component that is a fast-mode magnetoacoustic wave or shock wave and a slower apparent front formed as a result of successive stretching of closed magnetic field lines. In this scenario, the wave train and the successive loop expansion we observed likely correspond to the fast and slow components in the model, respectively.
\end{abstract}


\section{Introduction} \label{sec:intro}
Large-scale propagating disturbances in the solar corona were discovered by the Extreme-Ultraviolet Imaging Telescope (EIT) onboard the Solar and Heliospheric Observatory (SOHO), and termed ``EIT waves" or ``extreme-ultraviolet (EUV) waves" \citep{moses1997eit, thompson1998soho}. EUV waves propagating out from the eruption site at a speed of 50--1500 km s$^{-1}$ \citep{thompson2009catalog, nitta2013large}, and are usually accompanied by coronal mass ejections \citep[CMEs, e.g.,][]{biesecker2002solar, chen2009, nitta2013large, liu2018truly}. It was revealed recently that small-scale events, such as surges, jets, minifilaments, and mini-CMEs, may also drive EUV waves \citep[e.g.,][]{zheng2011possible, zheng2012homologous, shen2017small}). EUV waves are often found to propagate at nearly constant speeds or with decelerations (e.g., \citealt{warmuth2004multiwavelength, long2008kinematics}). In general, EUV waves are best seen as intensity enhancements in 193~\AA\ and 211~\AA\ passbands. A statistical investigation showed that among 138 events, 17\% appear as bright wavefronts and 41\% appear as dark wavefronts in the 171~\AA\ passband \citep{nitta2013large}.

Theoretical models of EUV waves can be classified into three groups: wave, non-wave, and hybrid models. Wave models interpret EUV waves as true waves, most commonly as fast-mode magnetohydrodynamic (MHD) waves or shocks \citep[e.g.,][]{thompson1999soho, Wang_2000, warmuth2001evolution}, and sometimes as slow-mode solitons \citep{wills2007eit}, magnetoacoustic surface gravity waves \citep{ballai2011magnetoacoustic}, or slow-mode waves and velocity vortices surrounding CMEs \citep{wang2009numerical}. The discovery of stationary EUV wavefront challenged the wave model \citep{delannee2000another}. Non-wave models interpret EUV waves as current shells between erupting magnetic field and the unperturbed magnetic fied \citep{delannee2000another, podladchikova2005automated, delannee2008new}, CME bubble projection \citep{2009AnGeo..27.3275A, ma2009new}, or successive reconnection fronts \citep{attrill2007coronal}. However, the observed reflections, refractions and transmissions of EUV waves \citep[e.g.,][]{gopal09, olmedo2012secondary, shen2013diffraction, zhou22} strongly favor the wave models. In particular, some observations showed the cospatiality of coronal EUV waves and chromospheric Moreton waves \citep[e.g.,][]{warmuth2001evolution, eto2002relation, okamoto2004filament, shen2012simultaneous, asai2012first, shen2019first, long2019quantifying, hou2022three}, which indicates some EUV waves are indeed fast-mode waves, rather than non-waves. To reconcile all these discrepancies, \citet{chen2002evidence, chen2005} proposed a hybrid model, i.e., there are two types of EUV waves, a faster component and a slower component. The fast-component EUV wave is a fast-mode MHD wave (or shock wave) whose footpoints would sweep the chromosphere so as to generate a Moreton wave, and the slow-component EUV wave is apparent propagation which is generated by successive magnetic field line stretching. Furthermore, as demonstrated and illustrated in Figures 3--4 of \citet{chen2002evidence}, each front of the slow-component EUV wave, as the magnetic field line stretches, is also a perturbation source, which would drive a fast-mode wave propagating ahead of the slow-component of EUV wave. That is to say, there should exist a wave train starting from the slow-component EUV wavefront but following the fast-component EUV wavefront, though their intensity is generally much fainter than the two components of EUV waves.

While the quasi-periodic wave train behind the fast-component EUV wave was clearly revealed in \citet{shen2019first} and \citet{zhou22}, the wave train starting from the slow-component EUV wave was discovered by \citet{Liu_2012}, although the authors claimed that the wave train starts from the CME frontal loop, without explicitly stating that the wave train starts from the slow-component EUV wave. Beside, it is still controversial about the physical process determining the period of the wave train. \citet{Liu_2012} and \citet{zhou22} found that the wave train shares the same period as the flare emission, whereas \citet{shen2019first} claimed that the period of the wave train is the same as that the untwisting motion of the erupting filament. Collecting more events is crucial in verifying the source of wave train and what determines the period of the wave train. In this letter, we report another EUV wave train event, which propagates across coronal loops. The observations and results are presented in Section \ref{sec:observations}, which are discussed in Section \ref{sec:disscussion}. Finally, we briefly summarize our results in Section \ref{sec:summary}.

\section{OBSERVATIONS AND RESULTS} \label{sec:observations}

On 2021 November 2, an M1.7-class flare, which peaked at $\sim$02:44 UT, occurred in NOAA active region (AR) 12891 at  N16$^\circ$E09$^\circ$ on the Sun. The flare was associated with a halo CME. An EUV wave event was detected in association with the flare. Among the ten channels of the Atmospheric Imaging Assembly (AIA; \citealt{2012SoPh..275...17L}) on board the Solar Dynamic Observatory (SDO), only the 171~\AA\ (dominated by emission from Fe\,{\sc{ix}}, sensitive to the $\sim$0.8 MK plasma) channel clearly reveals the multiple wavefronts, whereas the coronal loop stretching process can be seen in not only the 171~\AA\ channel but also other channels like 193~\AA\ (Fe\,{\sc{xii}}, $\sim$1.6 MK) and 211~\AA\ (Fe\,{\sc{xiv}}, $\sim$2.0 MK). These full-disk EUV images have a spatial resolution of $\sim$1.5$^{\prime\prime}$ and a cadence of 12 s.

\begin{figure}[ht!]
  \centering
  \includegraphics[width=1\textwidth,height=0.37\textwidth]{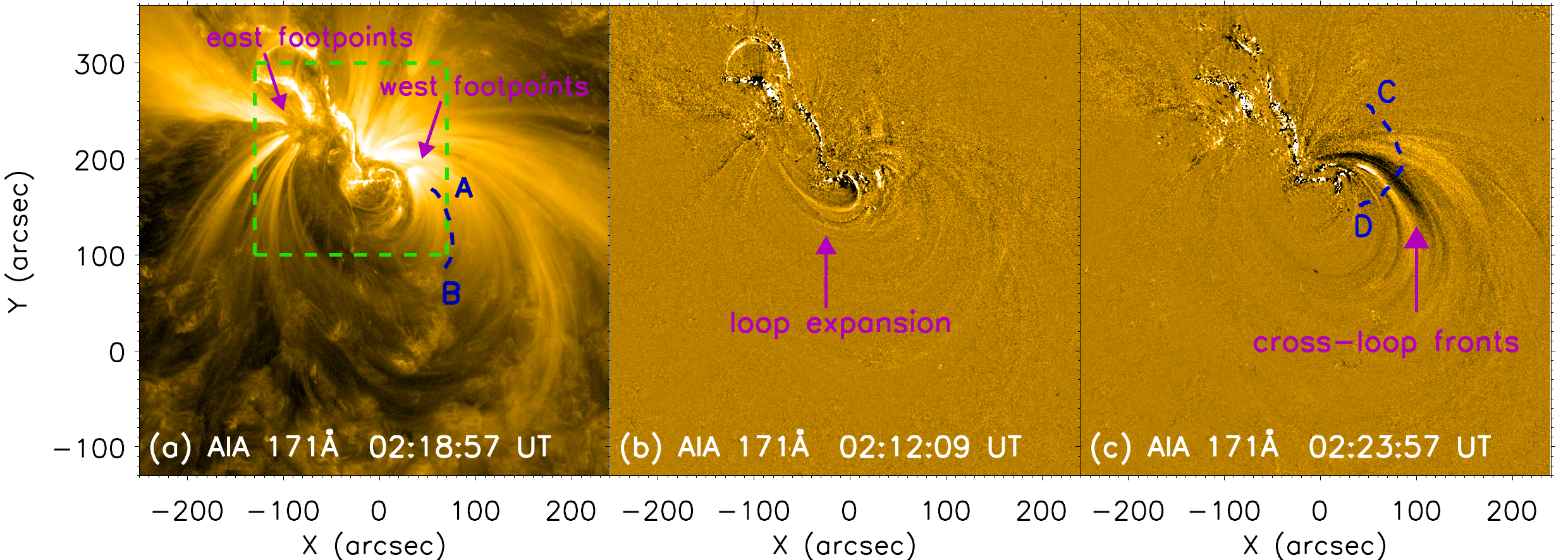}
  \caption{AIA 171~\AA\ image at 02:18:57 UT (a), and running difference images at 02:12:09 UT (b) and 02:23:57 UT (c). Slices A--B and C--D were used to obtain the time-distance diagrams shown in  Fig. \ref{fig:Fig.2}. Slice A--B was also chosen to obtain the intensity variations shown in  Fig. \ref{fig:Fig.3}. The dashed rectangle indicates the field of view of Fig. \ref{fig:Fig.4}(a).  The arrows in (b) and (c) indicate the loop expansion and the multiple wavefronts, respectively. (An animation of this figure is available online.)
  \label{fig:Fig.1}}
\end{figure}

Figure \ref{fig:Fig.1} presents an overview of the flare in the 171~\AA\ channel at 02:18:57 UT. During the early phase of the flare, we can see high-reaching overlying loops above the flare region. For the sake of convenience, the coronal loop footpoints located near the eastern flare ribbon are labeled as east footpoints, and the other as west footpoints. These overlying loops appear to be sequentially stretched up since $\sim$02:04 UT (see the online animation), resulting in an apparent ``EUV wave" phenomenon. A rough estimate yields a propagation speed of $\sim$95 km s$^{-1}$. Such a large velocity indicates that the loop expansion is more likely CME-driven rather than driven by magnetic flux emergence, which causes a speed of 20 km/s or smaller. Interestingly, AIA also observed another type of EUV wave in 171~\AA\ around the west footpoints of the coronal loops. Starting from $\sim$02:13 UT, we see quasi-periodic propagating disturbances with alternative bright and dark fronts across the loop legs in the running difference images (see the online animation), similar to those discovered by \citet{Liu_2012}. We can identify at least five distinct wavefronts sweeping across the loop legs around the west footpoints.

We utilized AIA 171~\AA\ running difference images ($\Delta t$=24 s) to study the propagation characteristics of the EUV waves. Figures \ref{fig:Fig.1}(b) and (c) display selected AIA 171~\AA\ running difference images. From panel (b), we can see that the fronts of loop expansion due to the loop stretching have a semicircular shape. From $\sim$02:04 UT to $\sim$02:18 UT, the associated disturbance propagates from the loop top to the loop legs around the west footpoints. The intensity enhancement is about 13\% relative to the background. When the disturbance arrives at the west footpoints at $\sim$02:13 UT, multiple cross-loop wavefronts appear. Multiple bright-dark interlaced fronts with a similar shape can be clearly identified from the running difference image sequence (see the online animation), and one pair of them is presented in Figure \ref{fig:Fig.1}(c). Their intensity variations are about 25\%, consistent with the typical intensity perturbations of EUV waves \citep[e.g.,][]{warmuth2015large}. The AIA observation also shows that the starting position of each wavefront is exactly at the location of each stretched loop leg around the west footpoints, which indicates that the origin of the multiple wavefronts is directly related to the loop stretching.

\begin{figure}
  \centering
  \plotone{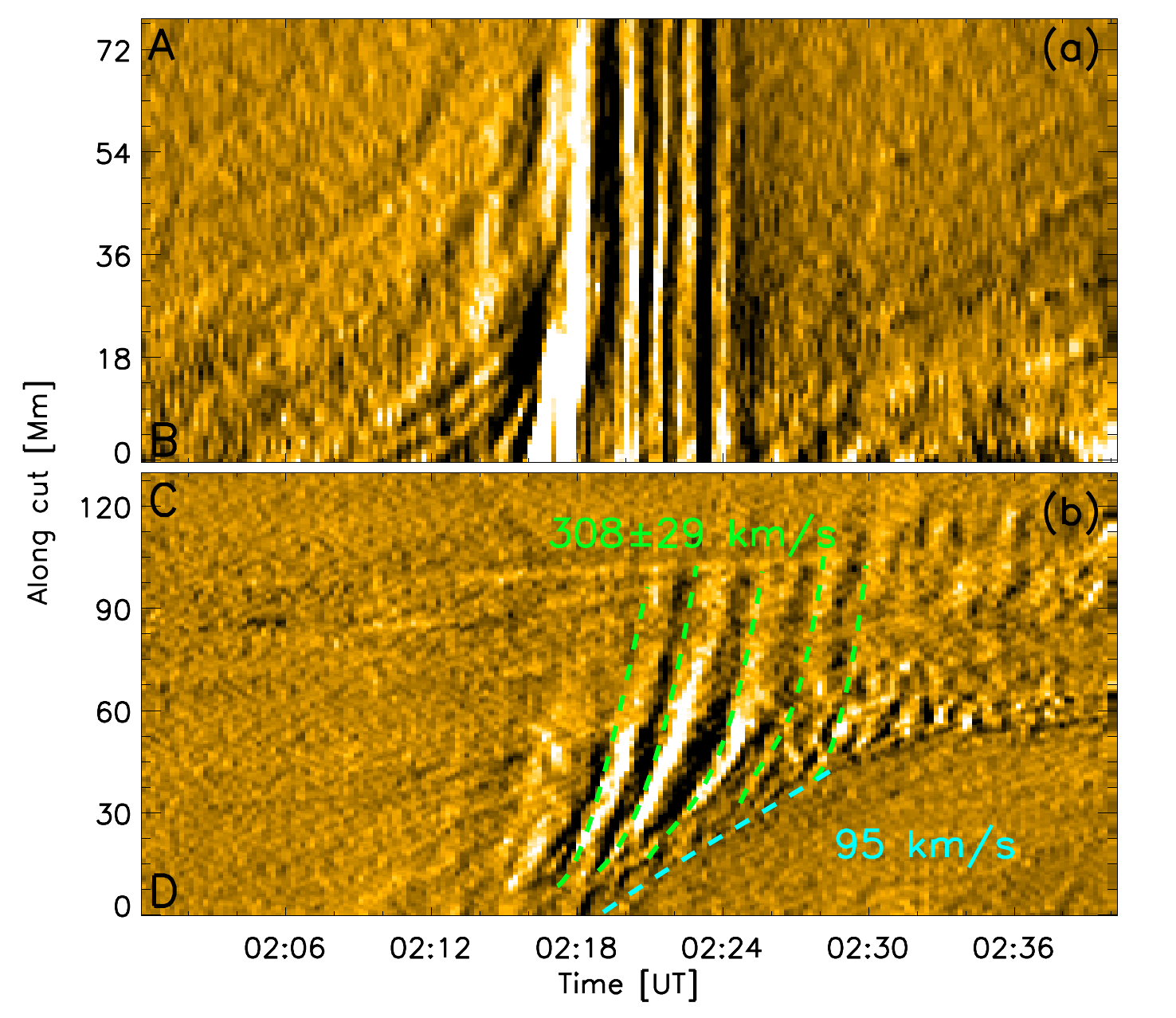}
  \caption{Time-distance diagrams of AIA 171~\AA\ running-difference images for the slice A--B (a) and slice C--D (b) in Fig. \ref{fig:Fig.1}.  The green and cyan dashed lines in (b) indicate the propagation of the multiple wavefronts and the envelope, respectively. Speeds of the propagating features are printed in the figure.
  \label{fig:Fig.2}}
\end{figure}

We selected the slices marked as the blue dashed lines in Figure \ref{fig:Fig.1}(a) and (c) to create time-distance diagrams, which are shown in Figure \ref{fig:Fig.2}. The vertical stripes in Figure \ref{fig:Fig.2}(a) indicate that there is no intensity disturbance propagating along the coronal loops when the multiple wavefronts were detected. This behavior suggests that the waves sweep across the loops, rather than propagate along the loops. Some weak inclined features before the appearance of these vertical stripes likely result from sweeping across the loops by the propagating waves. At leave five fast wavefronts with a similar morphology and speed can be clearly identified in Figure \ref{fig:Fig.2}(b). Such highly similar features indicate that they are likely homologous wave trains and have a high probability of being triggered by the same process. Each green line indicates a motion with a slight acceleration projected in the plane of sky. We thus applied a second-order polynomial fitting to each green line. The initial and final propagating speeds of these multiple wavefronts were found to be $280\pm 33$ km s$^{-1}$ and $330\pm 27$ km s$^{-1}$, with an average speed of $308\pm 29$ km s$^{-1}$. It should be noted that the speed is for the identified wave propagation across the loop legs rather than outwards away from the active region source. The period of the multiple wavefronts was found to be $\sim$120 s. All the starting points of these wavefronts in Figure \ref{fig:Fig.2}(b) form an envelope, which propagates away from the eruption site as indicated by the cyan dashed line. Such an apparent wave pattern continues to 02:38 UT even after the quasi-periodic wave train is not visible anymore. From the image sequence (see the online animation), we found that this envelope actually corresponds to the successively stretched bright loop legs around the west footpoints. We used a curved line to mark this envelope, which yielded an apparent speed of about $95\pm 8$ km s$^{-1}$. This means that the driving source of the quasi-periodic fast wavefronts moves at a speed of $\sim$95 km s$^{-1}$.

\begin{figure}
  \centering
  \plotone{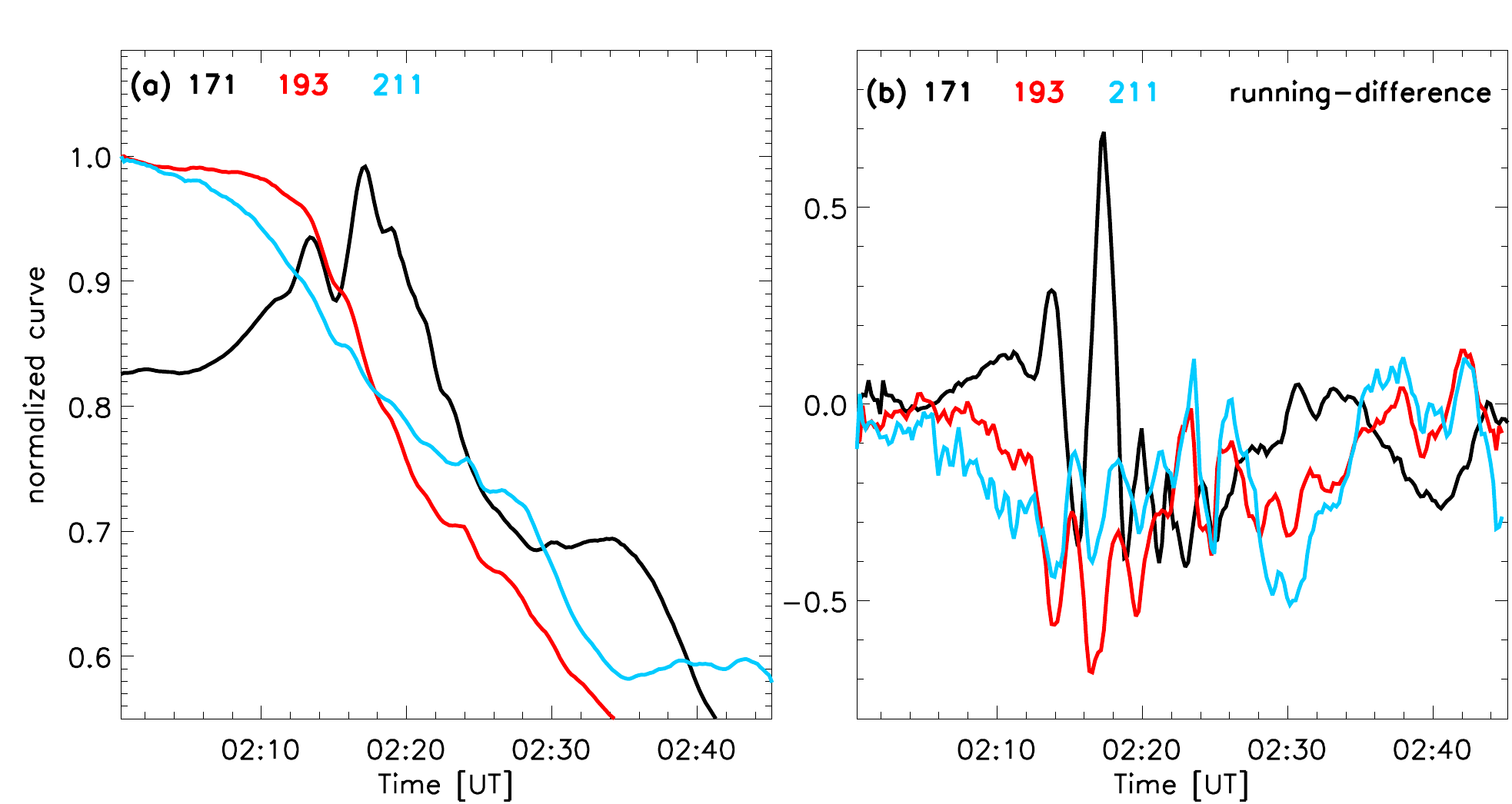}
  \caption{Variations of the intensities averaged over slice A--B in  Fig. \ref{fig:Fig.1}(a).
  Normalized intensity variations (a) and normalized running-difference of the intensity (b) in the 171~\AA, 193~\AA\ and 211~\AA\ channels (smoothed over 1-min interval).
  \label{fig:Fig.3}}
\end{figure}

As we mentioned above, the multiple fast wavefronts appear to be alternatively bright and dark in the 171~\AA\ running difference images, which have been seen in previous observations \citep[e.g.,][]{veronig2011plasma, li2012sdo, yang2013sdo, hou2022three}. The multiple wavefronts cannot be easily identified in other EUV channels. However, there are some weak wavefront-like perturbations in 193 \AA\ and 211 \AA\ near slice A--B in Figure \ref{fig:Fig.1}(a). Therefore, rather than trying to display quasi-periodic wave train in other channels, we picked up slice A--B where the wavefronts go through to make light curves in three EUV channels. Figure \ref{fig:Fig.3}(a) shows their normalized light curves. To eliminate the impact of the background evolution, we produced running-difference light curves, and the result is presented in Figure \ref{fig:Fig.3}(b). We can see significant oscillations in 171~\AA, which represent the alternatively bright and dark wavefronts. Similarly, the intensity oscillations are discernable in the 193 \AA\ and 211 \AA\ channels as well. In particular, we found that around the time when the intensity of 171~\AA\ reaches its minimum, the intensities of 193~\AA\ and 211~\AA\ reach their maxima. This anti-correlation indicates that the plasma at the wavefront is heated from $\lg (T/$K$) \approx 5.9$ to $\lg (T/$K$) \approx 6.2$, possibly due to compression (e.g., \citealt{wills1999observations, li2012sdo, hou2022three}).

To examine the magnetic field configuration of the flare region, we reconstructed the three-dimensional (3D) coronal magnetic field through a linear force-free field (LFFF) model described in \citet{1972SoPh...25..127N} and \citet{1981A&A...100..197A}. The extrapolation was based on the photospheric line-of-sight (LOS) magnetogram taken by the Magnetic Imager (HMI; \citealt{scherrer2012helioseismic}) on board SDO. The HMI magnetogram has a spatial resolution of 1$\arcsec$. We choose the time at $\sim$05:30 UT (after the flare) to obtain the position of relatively low post-flare loops. Using a trial-and-error method similar to \citet{2018SoPh..293...93C}, we found that an $\alpha$ value of $-2^{-7}$ Mm$^{-1}$ can achieve a good match between the extrapolated field lines and loop features in EUV images. Selected field lines are shown in Fig. \ref{fig:Fig.4}(a). We can see the high-reaching overlying loops above the low-lying post-flare loops, indicating that the stretching of the overlying coronal loops is presumably triggered by a rising filament (observed in AIA 304~\AA\, around 00:00 UT, not shown here) above the post-flare loops.

\section{DISSCUSSIONS} \label{sec:disscussion}

Multiple wavefronts are sometimes observed as quasi-periodic fast-mode propagating (QFP) waves in CME/flare eruptions. Some of them emanate from the flaring sites inside CMEs \citep[e.g.,][]{Liu_2011, Li_2018, Ofman_2018}. These QFP waves tend to be narrow in the angular span and often weak, e.g., with an intensity enhancement of about 1\%--8\% \citep{Liu_2011, shen2022coronal}. They are believed to be generated by the pressure pulses in flaring loops, propagating along the fan-like large-scale magnetic field. Correspondingly, these QFP waves have a period almost identical to that of the flare light curve. On the other hand, \citet{Liu_2012} discovered another type of QFP waves, which emanate from the CME frontal loop. These QFP wavefronts tend to be broader in the angular span and are brighter, e.g., with an intensity enhancement up to $>$20\%. The nature of these broad QFP wave train is still controversial.

Some authors claimed that these broad QFP waves originate from the flares \citep{Liu_2012, zhou22}. On the other hand, \citet{shen2019first} revealed that the QFP wave train immediately follows the fast-component EUV wave with the homogeneous shape. Since both slow-component and fast-component EUV waves are believed to be generated by coronal mass ejections \citep{chen2002evidence}, \citet{shen2019first} tended to support that the QFP waves are related to filament eruption. More importantly, they found that the QFP wave shares the common periodicity with the untwisting motion of the erupting filament, which reinforces the CME as the driving source. A similar conclusion was made by \citet{2019ApJ...871L...2M}, who found that the flare associated with a QFP wave train has no periodic pulsations. In this paper, we analyzed a QFP wave event associated with a halo CME and an M1.7-class flare. Unlike the narrow QFP waves channeled along open magnetic field with an EUV enhancement of 1--8\%, the QFP waves studied in this paper propagate across field lines, and their associated intensity enhancement is much larger. While the QFP waves propagate outward with a speed of $308\pm 29$ km s$^{-1}$, the starting positions of all the wavefronts form a wavelike pattern propagating out with a speed of 95 km s$^{-1}$. Such a wavelike pattern continues even after the QFP waves vanish. According to the hybrid model proposed by \citet{chen2002evidence, chen2005}, as a CME erupts, all the overlying magnetic field lines would be stretched up. Two types of EUV waves would be generated consequently: A piston-driven wave or shock wave propagates in the forefront, and an apparent wave propagates behind, which is generated via successive stretching of the magnetic field lines overlying the CME bubble. The two types of waves are often called fast-component and slow-component EUV waves. According to their model, the fast-component EUV wave, as a fast-mode MHD wave, should be $\sim$3 times faster than the slow-component EUV wave if the stretching magnetic field lines are concentric semicircles. In real observations, the magnetic field lines overlying the filament deviate from concentric semicircles, resulting the speed ratio in the range of $\sim$2--4. Furthermore, both their simulation results and the schematic sketch in \citet{chen2002evidence} pointed out that each front of the slow-component EUV wave is a source of perturbation, and would drive a fast-mode MHD wave, and the fast-component EUV wave or shock is simply the strongest leading front among the wave train. Our observations appear to fit into this hybrid model: The starting points of the QFP wavefronts form a wavelike pattern, corresponding to the slow-component EUV wave in the hybrid model, and the apparent speed of the slow-component EUV wave, i.e., 95 km s$^{-1}$, is roughly 3 times slower than the fast-mode MHD wave speed, $308\pm 29$ km s$^{-1}$. Similarly, the speed of the QFP waves in \citet{Liu_2012} is also 2--4 times faster than the lateral expanding velocity of the CME frontal loop. According to \citet{chen2009}, CME frontal loops are cospatial with the slow-component EUV wave. In this sense, the QFP wave speed is also 2--4 times faster than the slow-component EUV wave in \citet{Liu_2012}. This low speed, together with the weak intensity disturbance, suggests that the lateral expansion of the CME volume as often assumed in the ``wave" scenario \citep{patsourakos2012nature} is very weak. This is further seconded by the fact that no type II radio burst was observed in this event. In the absence of a strong lateral expansion, the \citet{chen2002evidence} scenario is favored \citep{Nitta2014Relation}. Therefore, we tend to believe that the broad QFP waves, as discovered by \citet{Liu_2012} and investigated in this paper, might be driven by the magnetic field line stretching as predicted by \citet{chen2002evidence, chen2005}. 

The period of QFP waves might also disclose the nature of QFP waves. \citet{Liu_2012} found that the QFP waves propagating northward have a period of 128 s, and the QFP waves propagating southward have a period of 212 s. These periods are similar to the periods of the flare pulsations, which are in the range of 2--3.5 min. Therefore, they concluded that the QFP waves originate from the flare site. The distinct periods of the northward and southward QFP waves provoke us to rethink the casual relationship between QFP waves and flares. It is more reasonable to see the QFP waves have the compound periods of the flares if the QFP waves originate from the flare site. The fact that the flare pulsations have the combined periods of the QFP waves reminds us of the opposite possibility that the QFP waves have a feedback to the erupting filament, which modulates the underlying magnetic reconnection in turn. The argument can be applied to the event in \citet{zhou22} as well. We checked the soft X-ray and EUV data in our event, and did not find any significant period that is close to the period of the quasi-periodic wave fronts. Therefore, the QFP wave train is not related to flare pulsations in this event. 

More importantly, we found in this paper that the bright QFP wavefronts always started from the bright legs of coronal loops. We can therefore envisage such a paradigm for the fast-mode MHD waves associated with filament eruptions according to the hybrid model of \citet{chen2002evidence}: If a CME eruption is in the background with relatively uniform magnetic and thermal structures, only the piston-driven shock wave would be seen \citep[e.g.,][]{thompson1998soho, liu2010first, veronig2011plasma, cheng2011investigation, yang2013sdo, kumar2015partial}. However, if the background magnetic and/or thermal structures are strongly non-uniform, e.g., those with discrete dense coronal loops, the stretching of a dense coronal loop would drive a fast-mode MHD wave with a distinct amplitude compared to the neighboring fast-mode MHD wave. As a result, a quasi-periodic wave train is formed. In this paradigm, the period of the QFP waves is actually determined by the separation of the dense coronal loops along the path of the QFP waves. 

\begin{figure}
  \centering
  \includegraphics[width=\textwidth]{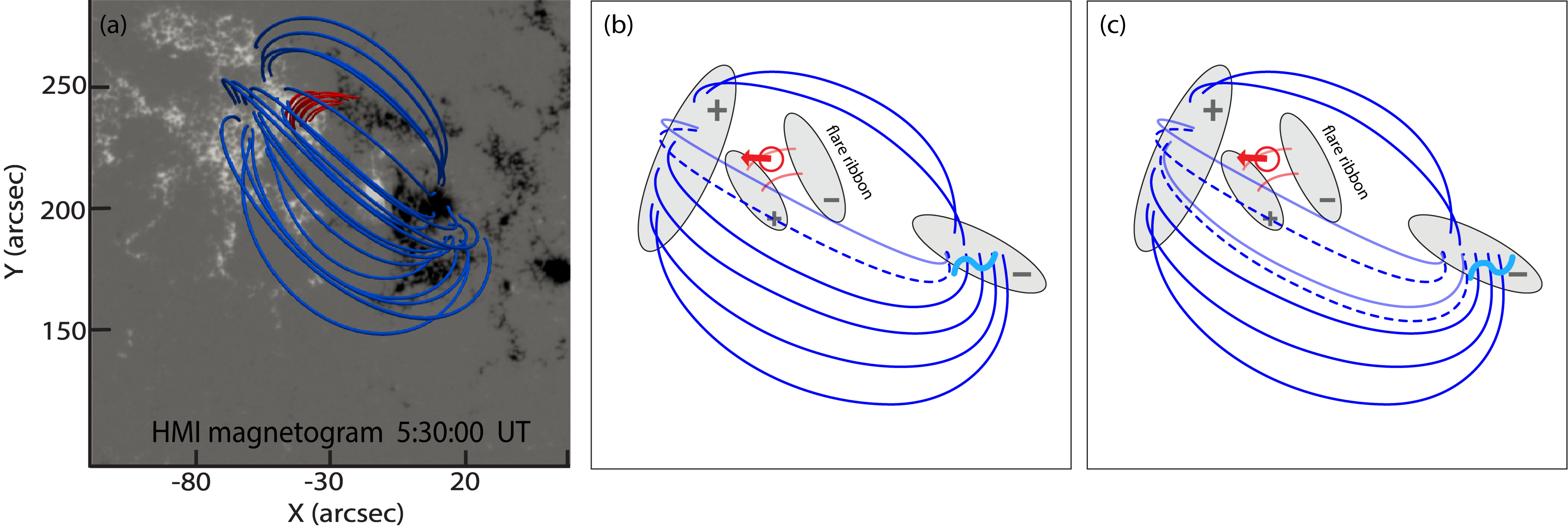}
  \caption{Magnetic field structure in the flare region. Panel (a) shows the linear force-free field extrapolation result of AR 12887 at $\sim$05:30 UT. The blue and red lines represent the overlying large-scale coronal loops and the post-flare loops, respectively. Panels (b) and (c) are skematic representations of the evolution of magnetic field structures. The plus and minus signs represent different magnetic polarities. The red thick lines between the positive and negative polarities correspond to the erupting filament. The dashed lines and semi-transparent solid lines represent the positions after and before the perturbation, respectively. The cyan wavy lines in Panel(b) and (c) indicate multiple wavefronts.
  \label{fig:Fig.4}}
\end{figure}

Based on the magnetic field extrapolation result shown in Figure \ref{fig:Fig.4}(a), we present a cartoon scenario to better explain the whole process in Figures \ref{fig:Fig.4}(b) and (c). During the impulsive phase of the flare, as the filament rises, the field lines overlying the filament are pushed to stretch up. The perturbation will propagate out with the phase velocity of fast-mode waves, while the deformation itself will be transferred to the outer field lines from the loop top to the legs. Each deformation or stretching in the west loop legs triggers a fast-mode wave. Therefore, the successive stretching of the magnetic field lines and the consequent fast-mode waves correspond to the non-wave slow-component and the fast-component EUV waves, respectively. The essence of this scenario is the same as that of \citet{chen2002evidence}. The update to their magnetic field line stretching model is that the background atmosphere is strongly inhomogeneous, with high-density plasma frozen in the blue solid lines. Once a field line with high density plasma is stretched up, a stronger fast-mode wavefront is produced.

The speed of the QFP wave in our paper, $\sim$308 km s$^{-1}$, appears to be at the lower end of typical coronal fast-mode MHD waves, and is also smaller than the speed of 775 km s$^{-1}$ in the model of \citet{chen2002evidence}. This relatively small speed probably cannot be explained by the projection effect, since the region of our event is close to the disk center. In this case the disturbances near the footpoints propagate almost along the surface of the Sun with a small projection effect, so the speeds of the multiple wavefronts should be close to the true fast-mode magnetoacoustic wave speed. The low value is probably due to the high plasma density since the waves were detected near the footpoints of the coronal loops. In fact, the density and magnetic field strength are strongly inhomogeneous in the corona, resulting in very different Alfv\'en speeds in different regions. The 308 km s$^{-1}$ speed is still within the normal coronal Alfv\'en speed range (e.g., \citealt{2020ScChE..63.2357Y,2020Sci...369..694Y}). The apparent propagating speed ($\sim$95 km s$^{-1}$) associated with the envelope in Figure \ref{fig:Fig.2}(b) is about one third of the fast wave speed, which is also consistent with the model of \citet{chen2002evidence}.

\section{SUMMARY} \label{sec:summary}

Several events of multiple EUV wavefronts have been previously reported \citep{liu2010first, Kumar_2017, shen2018homologous, zhou22}. However, most of these observations only revealed multi-front fast waves without a slow non-wave component. Similar to \citet{Liu_2012}, we observed an EUV wave event with multiple wavefronts, which is characterized by quasi-periodic wavefronts with a speed of $\sim$300 km s$^{-1}$ across legs of coronal loops. At least five wavefronts can be clearly identified and the period is $\sim$120 s. The origin sites of successive wavefronts are sequentially displaced, forming an apparent motion with a speed of $\sim$95 km s$^{-1}$. The multiple wavefronts are characterized by darkenings in AIA 171~\AA\ and brightenings in AIA 193 and 211~\AA, indicating plasma heating from $\lg (T/$K$) \approx 5.9$ to $\lg (T/$K$) \approx 6.2$ at the wavefront.

Our observational results might be explained by the theoretical model of \citet{chen2002evidence}, which implied multiple fast-mode wavefronts driven by the successive stretching of coronal loops during solar eruptions. The propagation of the slow-component EUV wave, which is non-wave in nature, is due to the apparent motion resulting from the successive loop stretching, whose speed is about one third of the fast wave speed as predicted by the model. According to the data analysis in this paper and their theoretical model, we propose that the period of the quasi-periodic waves is determined by the separation of the high-density coronal loops along the path of the wave propagation, rather than the flare pulsations.

\acknowledgments
  This work was supported by the National Key R\&D Program of China No. 2021YFA0718600 and NSFC grants 11825301, 12127901 and 42074204. AIA is a payload onboard \emph{SDO}, a mission of NASA's Living With a Star Program.

\bibliographystyle{aasjournal}
\bibliography{ref}

\end{document}